\documentclass[12pt]{article}
\usepackage{a4}
\usepackage{amssymb}
\usepackage{cite}
\newcommand{\be}{\begin{equation}}
\newcommand{\ee}{\end{equation}}
\newcommand{\bea}{\begin{eqnarray}}
\newcommand{\eea}{\end{eqnarray}}

\usepackage{hyperref}
\begin{document}
\thispagestyle{empty}
\def\thefootnote{\fnsymbol{footnote}}
\begin{center}\Large
Comments on fusion matrix in N=1
super Liouville field theory. \\
\vskip 2em
February 2016
\end{center}\vskip 0.2cm
\begin{center}
Hasmik Poghosyan$^{1}$
\footnote{hasmikpoghos@gmail.com}
 and Gor Sarkissian$^{1,2}$\footnote{ gor.sarkissian@ysu.am}
\end{center}
\vskip 0.2cm

\begin{center}
$^1$ Yerevan Physics Institute, \\
Alikhanian Br. 2, 0036\, Yerevan\\
Armenia
\end{center}
\begin{center}
$^2$Department of Physics, \ Yerevan State University,\\
Alex Manoogian 1, 0025\, Yerevan\\
Armenia
\end{center}

\vskip 1.5em
\begin{abstract} \noindent
We study several aspects of the $N=1$ super Liouville theory. We show that certain elements of the fusion matrix
in the Neveu-Schwarz sector related to the structure constants according to the same rules which we observe
in rational conformal field theory. We collect some evidences that these relations should hold also in the Ramond sector.
Using them
the Cardy-Lewellen equation for defects is studied, and defects are constructed.
\end{abstract}
\newpage
\tableofcontents
\newpage
\section{Introduction}
 During the last decades we got deep understanding of the properties of rational conformal field theories having
a finite number of primaries. Many important relations were obtained between basic notions of RCFT.
In particular we would like to mention the Verlinde formula \cite{Verlinde:1988sn}, relating matrix of modular transformation
and fusion coefficients, Moore-Seiberg relations between elements of fusion matrix, braiding matrix and matrix of modular transformations \cite{Moore:1988ss,Moore:1989vd,Moore:1988qv}.
We have formulas for boundary states \cite{Behrend:1999bn},
and defects \cite{Petkova:2000ip,Petkova:2001ag} in
rational conformal field theories.
Situation in non-rational conformal field theories is much more complicated. The infinite and even uncountable number
of primary fields is the main reason that progress in this direction is very slow.
One of the well studied non-rational theories is Liouville field theory.
Liouville field theory has attracted a lot of attention since Polyakov's suggestion to study strings in non-critical dimension.
Three-point correlation function (DOZZ formula) \cite{Dorn:1994xn,Zamolodchikov:1995aa} and fusing matrix \cite{Ponsot:1999uf} were found exactly. Other important examples of the non-rational CFT are
$N=1$ superconformal Liouville theory, conformal and superconformal Toda theories and more general para-Toda theories.
It is interesting to mention that all of them play a role in the recently established AGT correspondence
\cite{Alday:2009aq,Belavin:2011pp,Wyllard:2009hg,Nishioka:2011jk,Bershtein:2010wz,Bonelli:2011jx,Bonelli:2011kv,Bonelli:2012ny,Belavin:2012aa}.
Many data have been collected also in $N=1$ superconformal Liouville theory. In particular
three-point functions \cite{Poghosian:1996dw,Rashkov:1996np}
and the NS sector fusion matrices \cite{Hadasz:2007wi,Chorazkiewicz:2008es} have been found exactly.
Some attempts to find the fusion matrix also in the Ramond sector can be found in \cite{Chorazkiewicz:2011zd,Pawelkiewicz:2013wga}.
In this paper we study the following  relations, proved in rational CFT without multiplicities (fusion numbers $N^i_{jk}=0,1$), in $N=1$ super Liouville field theory:
\be\label{foifio}
F_{0,i}\left[\begin{array}{cc}
j&k\\
j&k^* \end{array}\right]F_{i,0}\left[\begin{array}{cc}
k^*&k\\
j&j \end{array}\right]={F_jF_k\over F_i}\, ,
\ee
where
\be
F_i\equiv F_{0,0}\left[\begin{array}{cc}
i&i^*\\
i&i \end{array}\right]={S_{00}\over S_{0i}}\, .
\ee
and
\be\label{cpijnx}
C^p_{ij}={\eta_i\eta_j\over \eta_0\eta_p} F_{0,p}\left[\begin{array}{cc}
j&i\\
j&i^* \end{array}\right]\, ,\quad \eta_i=\sqrt{C_{ii^*}/F_i}\, ,
\ee
which using (\ref{foifio}) can be written also
\be\label{cpij}
C^p_{ij}={\xi_i\xi_j\over \xi_0\xi_p}{1\over F_{p,0}\left[\begin{array}{cc}
j^*&j\\
i&i \end{array}\right]}\, ,\quad \xi_i=\eta_iF_i=\sqrt{C_{ii^*}F_i}\, .
\ee

The first relation (\ref{foifio}) is a consequence of the pentagon identity for fusion matrix \cite{Moore:1988ss,Moore:1989vd,Moore:1988qv}.
The second relation (\ref{cpijnx}) results from the bootstrap equation combined with the pentagon identity
\cite{Behrend:1999bn,Felder:1989hq,Fuchs:2004xi,Sarkissian:2011tr}.
These relations were examined in the Lioville field theory. The relation (\ref{foifio}) in the Lioville field theory was tested in \cite{Teschner:2008qh}. The relations (\ref{cpijnx}) and (\ref{cpij}) were examined in the Liouville field theory in \cite{Sarkissian:2011tr,Vartanov:2013ima}. In \cite{Sarkissian:2011tr} (\ref{cpijnx}) and (\ref{cpij}) in the Liouville field theory were
checked using the relation of the fusion matrix with boundary three-point function.
In \cite{Vartanov:2013ima} (\ref{cpijnx}) was checked using the following integral identity for the double Sine-functions $S_b(x)$:
\be\label{sintegg}
\int {dx\over i}\prod_{i=1}^3 S_b(x+a_i)
S_b(-x+b_i)=\prod_{i,j=1} S_b(a_i+b_j)\, ,
\ee
where
\be
\sum_i(a_i+b_i)=Q\, .
\ee
Recently it was found in \cite{Hadasz:2013bwa} supersymmetric generalization of this formula (eq.(\ref{sinteg} in text).
We find that also for $N=1$ super Liouville theory the susy version of this formula leads to the corresponding
generalization of the relations (\ref{cpijnx}) and (\ref{cpij}) in $N=1$ superLiouville theory.

The paper is organized as follows. In section \ref{qw1} we review basic facts on $N=1$ superLiouville theory.
In section \ref{qw2} we compute the elements of an Ansatz for the fusion matrices with
one of the intermediate states set to the vacuum.
In section \ref{qw5} we specialize the formulae obtained in section \ref{qw2} to the fusion matrices of the NS sector found in \cite{Hadasz:2007wi}.
In section \ref{qw6} we analyze the Ramond sector for a degenerate entry. In section \ref{qw7} we apply formulae obtained in section \ref{qw6}
to solve the Cardy-Lewellen equations for topological defects. In appendix some useful formulas are collected.

\section{N=1 Super Liouville field theory}
\label{qw1}
Let us review basic facts on the $N=1$ Super Liouville field theory.
$N=1$ super Liouville field theory is defined on a two-dimensional surface with metric $g_{ab}$ by the local Lagrangian
density
\be
{\cal L}={1\over 2\pi}g_{ab}\partial_a\varphi\partial_b \varphi+
{1\over 2\pi}(\psi\bar{\partial}\psi+\bar{\psi}\partial\bar{\psi})+2i\mu b^2\bar{\psi}\psi e^{b\varphi}+
2\pi \mu^2 b^2 e^{2b\varphi}\, ,
\ee
The energy-momentum tensor and the superconformal current are
\bea
&&T=-{1\over 2} (\partial\varphi\partial\varphi-Q\partial^2\varphi+\psi\partial\psi)\, ,\\
&&G=i(\psi\partial\varphi-Q\partial\psi)\, .
\eea

 The superconformal algebra is
 \bea
&& [L_m,L_n]=(m-n)L_{m+n}+{c\over 12}m(m^2-1)\delta_{m+n}\, ,\\
 &&[L_m, G_k]={m-2k\over 2}G_{m+k}\, ,\\
 &&\{G_k,G_l\}=2L_{l+k}+{c\over 3}\left(k^2-{1\over 4}\right)\delta_{k+l}\, ,
 \eea
with the central charge
\be
c_L={3\over 2}+3Q^2\, .
\ee
where
 \be
Q=b+{1\over b}\, .
\ee

Here $k$ and $l$ take integer values for the Ramond algebra and half-integer values for the Neveu-Schwarz algebra.

NS-NS primary fields $N_{\alpha}(z,\bar{z})$ in this theory,
$N_{\alpha}(z,\bar{z})=e^{\alpha \varphi(z,\bar{z})}$, have conformal dimensions
\be
\Delta^{NS}_{\alpha}={1\over 2}\alpha(Q-\alpha)\, .
\ee
The physical states have $\alpha={Q\over 2}+iP$.

Introduce also the field
\be
\tilde{N}_{\alpha}(z,\bar{z})=G_{-1/2}\bar{G}_{-1/2}N_{\alpha}(z,\bar{z})\, .
\ee

The R-R is defined as
\be
R_{\alpha}(z,\bar{z})=\sigma(z,\bar{z}) e^{\alpha \varphi(z,\bar{z})}\, ,
\ee
where $\sigma$ is the spin field \footnote{Sometimes  the Ramond field is defined as $R^{\pm}_{\alpha}(z,\bar{z})=\sigma^{\pm}(z,\bar{z}) e^{\alpha \varphi(z,\bar{z})}$, but in this paper the second field $R^{-}$ is not important.}.

The dimension of the R-R operator is
\be
\Delta_{\alpha}^R={1\over 16}+{1\over 2}\alpha(Q-\alpha)\, .
\ee

The NS-NS and R-R operators with the same conformal dimensions are proportional to each other, namely we have
\be
N_{\alpha}={\cal G}_{NS}(\alpha)N_{Q-\alpha}\, ,
\ee
\be
R_{\alpha}={\cal G}_R(\alpha)R_{Q-\alpha}\, ,
\ee
where ${\cal G}_{NS}(\alpha)$ and ${\cal G}_R(\alpha)$ are so called reflection functions. They also give two-point functions.
The elegant way to write the reflection functions is to introduce NS and R generalization of the
ZZ function \cite{Zamolodchikov:2001ah} in the bosonic Liouville theory:

\be\label{wn}
W_{NS}(\alpha)={2(\pi\mu\gamma(bQ/2))^{-{Q-2\alpha\over 2b}}\pi(\alpha-Q/2)\over \Gamma(1+b(\alpha-Q/2)) \Gamma(1+{1\over b}(\alpha-Q/2))}\, ,
\ee
\be\label{wr}
W_R(\alpha)={2\pi(\pi\mu\gamma(bQ/2))^{-{Q-2\alpha\over 2b}}\over \Gamma(1/2+b(\alpha-Q/2)) \Gamma(1/2+{1\over b}(\alpha-Q/2))}\, .
\ee
The reflection functions can be written
\be\label{dna}
{\cal G}_{NS}(\alpha)={W^{NS}(Q-\alpha)\over W^{NS}(\alpha)}\, ,
\ee
\be\label{dnr}
{\cal G}_R(\alpha)={W^R(Q-\alpha)\over W^R(\alpha)}\, .
\ee
The functions (\ref{wn}) and (\ref{wr}) satisfy also the relations
\be\label{wnmn}
W_{NS}(\alpha)W_{NS}(Q-\alpha)=-4\sin\pi b(\alpha-Q/2)\sin\pi {1\over b}(\alpha-Q/2)\, ,
\ee
\be\label{wrmr}
W_R(\alpha)W_R(Q-\alpha)=4\cos\pi b(\alpha-Q/2)\cos\pi {1\over b}(\alpha-Q/2)\, .
\ee
The degenerate states are given by the momenta:
\be\label{degnm}
\alpha_{m,n}={1\over 2b}(1-m)+{b\over 2}(1-n)
\ee
with even $m-n$ in the NS sector and odd $m-n$ in the R sector.

For the super conformal theory, characters are defined for the NS sector,
for the R sector and the $\widetilde{\rm NS}$ sector.
The corresponding characters for generic $P$ which have no null-states are
\be\label{charl}
\chi_{P}^{NS}(\tau)=\sqrt{\theta_3(q)\over \eta(q)}{q^{P^2/2}\over \eta(\tau)}\, ,
\ee
\be\label{char2}
\chi_{P}^{\widetilde{NS}}(\tau)=\sqrt{\theta_4(q)\over \eta(q)}{q^{P^2/2}\over \eta(\tau)}\, ,
\ee
\be\label{char3}
\chi_{P}^{R}(\tau)=\sqrt{\theta_2(q)\over 2\eta(q)}{q^{P^2/2}\over \eta(\tau)}\, ,
\ee
where $q=\exp(2\pi i\tau)$ and
\be
\eta(\tau)=q^{1/24}\prod_{n=1}^{\infty}(1-q^n)\, .
\ee
Modular transformation of characters (\ref{charl}) - (\ref{char3}) is well-known:
\be\label{motrp1}
\chi_{P}^{NS}(\tau)=\int\chi_{P'}^{NS}(-1/\tau)e^{-2i\pi PP'}dP'\, .
\ee
\be\label{motrp2}
\chi_{P}^{\widetilde{NS}}(\tau)=\int\chi_{P'}^R(-1/\tau)e^{-2i\pi PP'}dP'\, .
\ee
\be\label{motrp3}
\chi_{P}^R(\tau)=\int\chi_{P'}^{\widetilde{NS}}(-1/\tau)e^{-2i\pi PP'}dP'\, .
\ee

For degenerate representations, the characters are given by those of the corresponding Verma modules subtracted
by those of null submodules:
\be\label{deg1}
\chi_{m,n}^{NS}=\chi_{{1\over 2}(nb+mb^{-1})}^{NS}-\chi_{{1\over 2}(nb-mb^{-1})}^{NS}\, ,
\ee
\be\label{deg2}
\chi_{m,n}^{\widetilde{NS}}=\chi_{{1\over 2}(nb+mb^{-1})}^{\widetilde{NS}}-(-)^{rs}\chi_{{1\over 2}(nb-mb^{-1})}^{\widetilde{NS}}\, ,
\ee
\be\label{deg3}
\chi_{m,n}^{R}=\chi_{{1\over 2}(nb+mb^{-1})}^{R}-\chi_{{1\over 2}(nb-mb^{-1})}^{R}\, .
\ee
Modular transformations of (\ref{deg1}) - \ref{deg3}) are
\be\label{motrdi1}
\chi_{m,n}^{NS}(\tau)=\int\chi_{P}^{NS}(-1/\tau)2\sinh(\pi mP/b)\sinh(\pi nbP)dP\, .
\ee
\be\label{motrdi2}
\chi_{m,n}^{\widetilde{NS}}(\tau)=\int\chi_{P}^{R}(-1/\tau)2\sinh(\pi mP/b)\sinh(\pi nbP)dP\, , \quad m,n \quad {\rm even}\, ,
\ee
\be\label{motrdi3}
\chi_{m,n}^{\widetilde{NS}}(\tau)=\int\chi_{P}^{R}(-1/\tau)2\cosh(\pi mP/b)\cosh(\pi nbP)dP\, . \quad m,n \quad {\rm odd}\, .
\ee
Note that the vacuum component of the matrix of modular transformation specified by $(m,n)=(1,1)$
in formulae (\ref{motrdi1}) - (\ref{motrdi3}) coincide with the right hand side of  (\ref{wnmn}) and
(\ref{wrmr}) 

The structure constants in $N=1$ super Liouville field theory are computed in
\cite{Poghosian:1996dw,Rashkov:1996np}:
\bea\label{threep}
&&\langle N_{\alpha_1}(z_1,\bar{z}_1)N_{\alpha_2}(z_2,\bar{z}_2)N_{\alpha_3}(z_3,\bar{z}_3)\rangle=\\ \nonumber
&&{C_{NS}(\alpha_1,\alpha_2,\alpha_3)\over |z_{12}|^{2(\Delta_{\alpha_1}^N+\Delta_{\alpha_2}^N-\Delta_{\alpha_3}^N)}|z_{23}|^{2(\Delta_{\alpha_2}^N+\Delta_{\alpha_3}^N-
\Delta_{\alpha_1}^N)}|z_{13}|^{2(\Delta_{\alpha_1}^N+\Delta_{\alpha_3}^N-\Delta_{\alpha_2}^N)}}\, ,
\eea
\bea\label{threep}
&&\langle \tilde{N}_{\alpha_1}(z_1,\bar{z}_1)N_{\alpha_2}(z_2,\bar{z}_2)N_{\alpha_3}(z_3,\bar{z}_3)\rangle=\\ \nonumber
&&{\tilde{C}_{NS}(\alpha_1,\alpha_2,\alpha_3)\over |z_{12}|^{2(\Delta_{\alpha_1}^N+\Delta_{\alpha_2}^N-\Delta_{\alpha_3}^N+1/2)}|z_{23}|^{2(\Delta_{\alpha_2}^N+\Delta_{\alpha_3}^N-
\Delta_{\alpha_1}^N-1/2)}|z_{13}|^{2(\Delta_{\alpha_1}^N+\Delta_{\alpha_3}^N-\Delta_{\alpha_2}^N+1/2)}
 }\, ,
\eea
\bea\label{threep}
&&\langle R_{\alpha_1}(z_1,\bar{z}_1)R_{\alpha_2}(z_2,\bar{z}_2)N_{\alpha_3}(z_3,\bar{z}_3)\rangle=\\ \nonumber
&&{C_{R}(\alpha_1,\alpha_2|\alpha_3)+\tilde{C}_{R}(\alpha_1,\alpha_2|\alpha_3)\over |z_{12}|^{2(\Delta_{\alpha_1}^R+\Delta_{\alpha_2}^R-\Delta_{\alpha_3}^N)}|z_{23}|^{2(\Delta_{\alpha_2}^R+\Delta_{\alpha_3}^N-
\Delta_{\alpha_1}^R)}|z_{13}|^{2(\Delta_{\alpha_1}^R+\Delta_{\alpha_3}^N-\Delta_{\alpha_2}^R)}
 }\, ,
\eea
where $z_{ij}=z_i-z_j$,

and

\bea\label{dozz1}
&&C_{NS}(\alpha_1,\alpha_2,\alpha_3)=\lambda^{(Q-\sum_{i=1}^3\alpha_i)/b}\times\\ \nonumber
&&{\Upsilon'_{NS}(0)\Upsilon_{NS}(2\alpha_1)\Upsilon_{NS}(2\alpha_2)\Upsilon_{NS}(2\alpha_3)\over
\Upsilon_{NS}(\alpha_1+\alpha_2+\alpha_3-Q)\Upsilon_{NS}(\alpha_1+\alpha_2-\alpha_3)
\Upsilon_{NS}(\alpha_2+\alpha_3-\alpha_1)\Upsilon_{NS}(\alpha_3+\alpha_1-\alpha_2)}\, ,
\eea
\bea\label{dozz2}
&&\tilde{C}_{NS}(\alpha_1,\alpha_2,\alpha_3)=\lambda^{(Q-\sum_{i=1}^3\alpha_i)/b}\times\\ \nonumber
&&{\Upsilon'_{NS}(0)\Upsilon_{NS}(2\alpha_1)\Upsilon_{NS}(2\alpha_2)\Upsilon_{NS}(2\alpha_3)\over
\Upsilon_R(\alpha_1+\alpha_2+\alpha_3-Q)\Upsilon_R(\alpha_1+\alpha_2-\alpha_3)
\Upsilon_R(\alpha_2+\alpha_3-\alpha_1)\Upsilon_R(\alpha_3+\alpha_1-\alpha_2)}\, ,
\eea

\bea\label{dozz1r}
&&C_{R}(\alpha_1,\alpha_2|\alpha_3)=\lambda^{(Q-\sum_{i=1}^3\alpha_i)/b}\times\\ \nonumber
&&{\Upsilon'_{NS}(0)\Upsilon_{R}(2\alpha_1)\Upsilon_{R}(2\alpha_2)\Upsilon_{NS}(2\alpha_3)\over
\Upsilon_{R}(\alpha_1+\alpha_2+\alpha_3-Q)\Upsilon_{R}(\alpha_1+\alpha_2-\alpha_3)
\Upsilon_{NS}(\alpha_2+\alpha_3-\alpha_1)\Upsilon_{NS}(\alpha_3+\alpha_1-\alpha_2)}\, ,
\eea
\bea\label{dozz2r}
&&\tilde{C}_{R}(\alpha_1,\alpha_2|\alpha_3)=\lambda^{(Q-\sum_{i=1}^3\alpha_i)/b}\times\\ \nonumber
&&{\Upsilon'_{NS}(0)\Upsilon_{R}(2\alpha_1)\Upsilon_{R}(2\alpha_2)\Upsilon_{NS}(2\alpha_3)\over
\Upsilon_{NS}(\alpha_1+\alpha_2+\alpha_3-Q)\Upsilon_{NS}(\alpha_1+\alpha_2-\alpha_3)
\Upsilon_R(\alpha_2+\alpha_3-\alpha_1)\Upsilon_R(\alpha_3+\alpha_1-\alpha_2)}\, ,
\eea
and
\be
\lambda=\pi\mu\gamma\left({bQ\over 2}\right)b^{1-b^2}\, .
\ee
Fusion matrix in the NS sector is computed in \cite{Hadasz:2007wi,Chorazkiewicz:2008es}.
Let us denote
\be
F_{\alpha_s,\alpha_t}\left[\begin{array}{cc}
\alpha_3& \alpha_2\\
\alpha_4& \alpha_1 \end{array}\right]^1_1\equiv F_{N_{\alpha_s},N_{\alpha_t}}\left[\begin{array}{cc}
N_{\alpha_3}& N_{\alpha_2}\\
N_{\alpha_4}& N_{\alpha_1} \end{array}\right]\, ,\hspace{0.3cm}
F_{\alpha_s,\alpha_t}\left[\begin{array}{cc}
\alpha_3& \alpha_2\\
\alpha_4& \alpha_1 \end{array}\right]^2_1\equiv F_{N_{\alpha_s},\tilde{N}_{\alpha_t}}\left[\begin{array}{cc}
N_{\alpha_3}& N_{\alpha_2}\\
N_{\alpha_4}& N_{\alpha_1} \end{array}\right]\, ,
\ee
\be
F_{\alpha_s,\alpha_t}\left[\begin{array}{cc}
\alpha_3& \alpha_2\\
\alpha_4& \alpha_1 \end{array}\right]^1_2\equiv F_{\tilde{N}_{\alpha_s},N_{\alpha_t}}\left[\begin{array}{cc}
N_{\alpha_3}& N_{\alpha_2}\\
N_{\alpha_4}& N_{\alpha_1} \end{array}\right]\, ,\hspace{0.3cm}
F_{\alpha_s,\alpha_t}\left[\begin{array}{cc}
\alpha_3& \alpha_2\\
\alpha_4& \alpha_1 \end{array}\right]^2_2\equiv F_{\tilde{N}_{\alpha_s},\tilde{N}_{\alpha_t}}\left[\begin{array}{cc}
N_{\alpha_3}& N_{\alpha_2}\\
N_{\alpha_4}& N_{\alpha_1} \end{array}\right]\, .
\ee

To write the fusion matrix we use the following convention. The functions $\Upsilon_i, \Gamma_i, S_i$ will
be understood $\Upsilon_{NS}, \Gamma_{NS}, S_{NS}$ for $i=1\;\rm{mod}\; 2$, and $\Upsilon_R, \Gamma_R, S_R$
for $i=0\; \rm{mod}\; 2$. Now we can write the fusion matrix:
 \bea\label{fusja}
&&F_{\alpha_s,\alpha_t}\left[\begin{array}{cc}
\alpha_3& \alpha_2\\
\alpha_4& \alpha_1 \end{array}\right]^i_j=\\ \nonumber
&&{\Gamma_i(2Q-\alpha_t-\alpha_2-\alpha_3)\Gamma_i(Q-\alpha_t+\alpha_3-\alpha_2)\Gamma_i(Q+\alpha_t-\alpha_2-\alpha_3)
\Gamma_i(\alpha_3+\alpha_t-\alpha_2)\over
\Gamma_j(2Q-\alpha_1-\alpha_s-\alpha_2)\Gamma_{j}(Q-\alpha_s-\alpha_2+\alpha_1)\Gamma_j(Q-\alpha_1-\alpha_2+\alpha_s)
\Gamma_{j}(\alpha_s+\alpha_1-\alpha_2)}\\  \nonumber
&\times&{\Gamma_i(Q-\alpha_t-\alpha_1+\alpha_4)\Gamma_i(\alpha_1+\alpha_4-\alpha_t)\Gamma_i(\alpha_t+\alpha_4-\alpha_1)
\Gamma_i(\alpha_t+\alpha_1+\alpha_4-Q)\over
\Gamma_{j}(Q-\alpha_s-\alpha_3+\alpha_4)\Gamma_j(\alpha_3+\alpha_4-\alpha_s)\Gamma_{j}(\alpha_s+\alpha_4-\alpha_3)
\Gamma_j(\alpha_s+\alpha_3+\alpha_4-Q)}\\  \nonumber
&&\times {\Gamma_{\rm NS}(2Q-2\alpha_s)\Gamma_{\rm NS}(2\alpha_s)\over \Gamma_{\rm NS}(Q-2\alpha_t)\Gamma_{\rm NS}(2\alpha_t-Q)}
{1\over i}\int_{-i\infty}^{i\infty}d\tau J_{\alpha_s,\alpha_t}\left[\begin{array}{cc}
\alpha_3& \alpha_2\\
\alpha_4& \alpha_1 \end{array}\right]^i_j\, ,
\eea

\bea
&&J_{\alpha_s,\alpha_t}\left[\begin{array}{cc}
\alpha_3& \alpha_2\\
\alpha_4& \alpha_1 \end{array}\right]^1_1=\\ \nonumber
&& {S_{NS}(Q+\tau-\alpha_1)S_{\rm NS}(\tau+\alpha_4+\alpha_2-\alpha_3)S_{NS}(\tau+\alpha_1)S_{NS}(\tau+\alpha_4+\alpha_2+\alpha_3-Q)\over
S_{NS}(Q+\tau+\alpha_4-\alpha_t)S_{NS}(\tau+\alpha_4+\alpha_t)S_{NS}(Q+\tau+\alpha_2-\alpha_s)S_{NS}(\tau+\alpha_2+\alpha_s)}\\ \nonumber
&+&{S_{R}(Q+\tau-\alpha_1)S_{R}(\tau+\alpha_4+\alpha_2-\alpha_3)S_{R}(\tau+\alpha_1)S_{R}(\tau+\alpha_4+\alpha_2+\alpha_3-Q)\over
S_{R}(Q+\tau+\alpha_4-\alpha_t)S_{R}(\tau+\alpha_4+\alpha_t)S_{R}(Q+\tau+\alpha_2-\alpha_s)S_{R}(\tau+\alpha_2+\alpha_s)}\, ,
\eea

\bea
&&J_{\alpha_s,\alpha_t}\left[\begin{array}{cc}
\alpha_3& \alpha_2\\
\alpha_4& \alpha_1 \end{array}\right]^1_2=\\ \nonumber
&& {S_{NS}(Q+\tau-\alpha_1)S_{\rm NS}(\tau+\alpha_4+\alpha_2-\alpha_3)S_{NS}(\tau+\alpha_1)S_{NS}(\tau+\alpha_4+\alpha_2+\alpha_3-Q)\over
S_{NS}(Q+\tau+\alpha_4-\alpha_t)S_{NS}(\tau+\alpha_4+\alpha_t)S_{R}(Q+\tau+\alpha_2-\alpha_s)S_{R}(\tau+\alpha_2+\alpha_s)}\\ \nonumber
&-&{S_{R}(Q+\tau-\alpha_1)S_{R}(\tau+\alpha_4+\alpha_2-\alpha_3)S_{R}(\tau+\alpha_1)S_{R}(\tau+\alpha_4+\alpha_2+\alpha_3-Q)\over
S_{R}(Q+\tau+\alpha_4-\alpha_t)S_{R}(\tau+\alpha_4+\alpha_t)S_{NS}(Q+\tau+\alpha_2-\alpha_s)S_{NS}(\tau+\alpha_2+\alpha_s)}\, ,
\eea

\bea
&&J_{\alpha_s,\alpha_t}\left[\begin{array}{cc}
\alpha_3& \alpha_2\\
\alpha_4& \alpha_1 \end{array}\right]^2_1=\\ \nonumber
&& {S_{NS}(Q+\tau-\alpha_1)S_{\rm NS}(\tau+\alpha_4+\alpha_2-\alpha_3)S_{NS}(\tau+\alpha_1)S_{NS}(\tau+\alpha_4+\alpha_2+\alpha_3-Q)\over
S_{R}(Q+\tau+\alpha_4-\alpha_t)S_{R}(\tau+\alpha_4+\alpha_t)S_{NS}(Q+\tau+\alpha_2-\alpha_s)S_{NS}(\tau+\alpha_2+\alpha_s)}\\ \nonumber
&-&{S_{R}(Q+\tau-\alpha_1)S_{R}(\tau+\alpha_4+\alpha_2-\alpha_3)S_{R}(\tau+\alpha_1)S_{R}(\tau+\alpha_4+\alpha_2+\alpha_3-Q)\over
S_{NS}(Q+\tau+\alpha_4-\alpha_t)S_{NS}(\tau+\alpha_4+\alpha_t)S_{R}(Q+\tau+\alpha_2-\alpha_s)S_{R}(\tau+\alpha_2+\alpha_s)}\, ,
\eea

\bea
&&J_{\alpha_s,\alpha_t}\left[\begin{array}{cc}
\alpha_3& \alpha_2\\
\alpha_4& \alpha_1 \end{array}\right]^2_2=\\ \nonumber
&& {S_{NS}(Q+\tau-\alpha_1)S_{\rm NS}(\tau+\alpha_4+\alpha_2-\alpha_3)S_{NS}(\tau+\alpha_1)S_{NS}(\tau+\alpha_4+\alpha_2+\alpha_3-Q)\over
S_{R}(Q+\tau+\alpha_4-\alpha_t)S_{R}(\tau+\alpha_4+\alpha_t)S_{R}(Q+\tau+\alpha_2-\alpha_s)S_{R}(\tau+\alpha_2+\alpha_s)}\\ \nonumber
&+&{S_{R}(Q+\tau-\alpha_1)S_{R}(\tau+\alpha_4+\alpha_2-\alpha_3)S_{R}(\tau+\alpha_1)S_{R}(\tau+\alpha_4+\alpha_2+\alpha_3-Q)\over
S_{NS}(Q+\tau+\alpha_4-\alpha_t)S_{NS}(\tau+\alpha_4+\alpha_t)S_{NS}(Q+\tau+\alpha_2-\alpha_s)S_{NS}(\tau+\alpha_2+\alpha_s)}\, .
\eea

\section{Values of the fusion matrix for the intermediate vacuum states}
\label{qw2}
\subsection{$\alpha_s\to 0$}
\label{qw3}
Motivated by the form of structure constants (\ref{dozz1})-(\ref{dozz2r}) and fusing matrix
(\ref{fusja}) we define the following general expressions for the fusion matrix:
\be\label{fmnr}
F^{\cal I}_{\alpha_s,\alpha_t}\left[\begin{array}{cc}
\alpha_3& \alpha_2\\
\alpha_4& \alpha_1 \end{array}\right]=
{M^{\cal I}\over i}\int_{-i\infty}^{i\infty}d\tau J^{\cal I}_{\alpha_s,\alpha_t}\left[\begin{array}{cc}
\alpha_3& \alpha_2\\
\alpha_4& \alpha_1 \end{array}\right]
\ee
with
\bea
&&M^{\cal I}=\\ \nonumber
&&{\Gamma_A(2Q-\alpha_t-\alpha_2-\alpha_3)\Gamma_B(Q-\alpha_t+\alpha_3-\alpha_2)\Gamma_C(Q+\alpha_t-\alpha_2-\alpha_3)
\Gamma_D(\alpha_3+\alpha_t-\alpha_2)\over
\Gamma_E(2Q-\alpha_1-\alpha_s-\alpha_2)\Gamma_{NS}(Q-\alpha_s-\alpha_2+\alpha_1)\Gamma_E(Q-\alpha_1-\alpha_2+\alpha_s)
\Gamma_{NS}(\alpha_s+\alpha_1-\alpha_2)}\\  \nonumber
&\times&{\Gamma_B(Q-\alpha_t-\alpha_1+\alpha_4)\Gamma_C(\alpha_1+\alpha_4-\alpha_t)\Gamma_D(\alpha_t+\alpha_4-\alpha_1)
\Gamma_A(\alpha_t+\alpha_1+\alpha_4-Q)\over
\Gamma_{NS}(Q-\alpha_s-\alpha_3+\alpha_4)\Gamma_F(\alpha_3+\alpha_4-\alpha_s)\Gamma_{NS}(\alpha_s+\alpha_4-\alpha_3)
\Gamma_F(\alpha_s+\alpha_3+\alpha_4-Q)}\\  \nonumber
&&\times {\Gamma_{\rm NS}(2Q-2\alpha_s)\Gamma_{\rm NS}(2\alpha_s)\over \Gamma_{\rm L}(Q-2\alpha_t)\Gamma_{\rm L}(2\alpha_t-Q)}\, ,
\eea
\bea\label{jcali}
&&J^{\cal I}_{\alpha_s,\alpha_t}\left[\begin{array}{cc}
\alpha_3& \alpha_2\\
\alpha_4& \alpha_1 \end{array}\right]=\\ \nonumber
&& {S_{\nu_1}(Q+\tau-\alpha_1)S_{\rm K}(\tau+\alpha_4+\alpha_2-\alpha_3)S_{\nu_2}(\tau+\alpha_1)S_{\nu_3}(\tau+\alpha_4+\alpha_2+\alpha_3-Q)\over
S_{\mu_1+1}(Q+\tau+\alpha_4-\alpha_t)S_{\mu_2+1}(\tau+\alpha_4+\alpha_t)S_{\mu_3+1}(Q+\tau+\alpha_2-\alpha_s)S_{\rm K}(\tau+\alpha_2+\alpha_s)}\\ \nonumber
&+&\eta{S_{\nu_1+1}(Q+\tau-\alpha_1)S_{\rm K+1}(\tau+\alpha_4+\alpha_2-\alpha_3)S_{\nu_2+1}(\tau+\alpha_1)S_{\nu_3+1}(\tau+\alpha_4+\alpha_2+\alpha_3-Q)\over
S_{\mu_1}(Q+\tau+\alpha_4-\alpha_t)S_{\mu_2}(\tau+\alpha_4+\alpha_t)S_{\mu_3}(Q+\tau+\alpha_2-\alpha_s)S_{{\rm K}+1}(\tau+\alpha_2+\alpha_s)}\, ,
\eea
where $\eta=(-1)^{ (1+\sum_i (\nu_i+\mu_i))/2}$.
${\cal I}$ denotes fusion matrices of  different structures, and capital Latin
letters here take values $NS$ and $R$. The expressions similar to (\ref{jcali}) were considered also in \cite{Pawelkiewicz:2013wga} 
in construction of the Racah-Wigner coefficients. 

Define also the following general expression for structure constants:

\bea\label{dozz}
&&C_{\cal I}(\alpha_1,\alpha_2,\alpha_3)=\lambda^{(Q-\sum_{i=1}^3\alpha_i)/b}\times\\ \nonumber
&&{\Upsilon'_{NS}(0)\Upsilon_L(2\alpha_1)\Upsilon_E(2\alpha_2)\Upsilon_F(2\alpha_3)\over
\Upsilon_A(\alpha_1+\alpha_2+\alpha_3-Q)\Upsilon_B(\alpha_1+\alpha_2-\alpha_3)
\Upsilon_C(\alpha_2+\alpha_3-\alpha_1)\Upsilon_D(\alpha_3+\alpha_1-\alpha_2)}\, ,
\eea

Now consider the limit:
\be\label{lim-s}
\alpha_s=\epsilon\to 0,\hspace{1cm} \alpha_3=\alpha_4, \hspace{1cm}\alpha_1=\alpha_2\, .
\ee
In this limit using formulae from appendix and the definition (\ref{dozz}) we get for the factor in front of integral:

\bea\label{calph2}
M^{\cal I}\to C_{\cal I}(\alpha_t,\alpha_1,\alpha_3){W_{NS}(Q)W_F(\alpha_3)W_L(\alpha_t)
\over 2\pi W_E(Q-\alpha_1)}\times
\\ \nonumber
{S_B(Q-\alpha_t+\alpha_3-\alpha_1)S_D(\alpha_3+\alpha_t-\alpha_1)
S_{\rm E}(2\alpha_1)\over  S_{\rm F}(2\alpha_3)S_{NS}(\epsilon)}\, .
\eea

Let us now evaluate the integral part of (\ref{fmnr}) in the limit (\ref{lim-s}).
For this purpose we will use the
 formula \cite{Hadasz:2013bwa}
\be\label{sinteg}
\sum_{\nu =0,1}(-1)^{\nu (1+\sum_i (\nu_i+\mu_i))/2}\int {dx\over i}\prod_{i=1}^3 S_{\nu+\nu_i}(x+a_i)
S_{1+\nu+\mu_i}(-x+b_i)=2\prod_{i,j=1} S_{\nu_i+\mu_j}(a_i+b_j)\, ,
\ee
\be\label{vmn2}
\sum_i(\nu_i+\mu_i)=1\; {\rm mod}\; 2\, ,
\ee
and
\be
\sum_i(a_i+b_i)=Q\, .
\ee

First note that in the limit (\ref{lim-s}) the arguments of $S_K$'s in numerator and denominator  coincide and they get canceled.

For the rest of $S$'s in this limit we get for  $a_i$ in the argument of $S_{\nu_i}(\tau+a_i)$ and $b_i$ in the argument of $S_{\mu_i+1}(-\tau+b_i)$:
\bea\label{qabs}
&&a_1=Q-\alpha_1\, .\hspace{2.1cm} b_1=\alpha_t-\alpha_3\, ,\\ \nonumber
&&a_2=\alpha_1\, ,\hspace{2.9cm} b_2=Q-\alpha_3-\alpha_t\, ,\\ \nonumber
&&a_3=2\alpha_3+\alpha_1-Q\, ,\hspace{1cm} b_3=-\alpha_1\, .\\ \nonumber
\eea
From (\ref{qabs}) we obtain
\bea\label{sa1b1}
a_1+b_1=Q-\alpha_1+\alpha_t-\alpha_3\, ,\\ \nonumber
a_1+b_2=2Q-\alpha_1-\alpha_3-\alpha_t\, ,\\ \nonumber
a_1+b_3=Q-2\alpha_1\, ,
\eea
\bea\label{sa2b2}
a_2+b_1=\alpha_1+\alpha_t-\alpha_3\, ,\\ \nonumber
a_2+b_2=Q+\alpha_1-\alpha_3-\alpha_t\, ,\\ \nonumber
a_2+b_3=\epsilon\, ,
\eea
\bea\label{sa3b3}
a_3+b_1=\alpha_3+\alpha_t+\alpha_1-Q\, ,\\ \nonumber
a_3+b_2=\alpha_1+\alpha_3-\alpha_t\, ,\\ \nonumber
a_3+b_3=2\alpha_3-Q\, .
\eea
Note that
\bea
a_1+b_1=Q-(a_3+b_2)\, ,\\ \nonumber
a_1+b_2=Q-(a_3+b_1)\, ,
\eea
and
\be\label{balq}
\sum_i(a_i+b_i)=Q\, .
\ee
Let us impose also
\bea\label{vmn1}
&&\nu_1+\mu_1=\nu_3+\mu_2\quad \rm{mod}\; 2\, ,\\ \nonumber
&&\nu_1+\mu_2=\nu_3+\mu_1\quad \rm{mod}\; 2\, ,\\ \nonumber
&&\nu_2+\mu_3=1\quad \rm{mod}\; 2\, .
\eea

Assuming also that (\ref{vmn2}) is satisfied we get from (\ref{sinteg})  using formulas (\ref{sa1b1})-(\ref{vmn1})

\be
{1\over i}\int_{-i\infty}^{i\infty}d\tau J^{\cal I}_{\alpha_s,\alpha_t}\left[\begin{array}{cc}
\alpha_3& \alpha_2\\
\alpha_4& \alpha_1 \end{array}\right]\to{2S_{\nu_2+\mu_1}(\alpha_1+\alpha_t-\alpha_3)
S_{\nu_3+\mu_3}(2\alpha_3-Q)S_{\rm NS}(\epsilon)
\over S_{\nu_1+\mu_3}(2\alpha_1)S_{\nu_2+\mu_2}(\alpha_3+\alpha_t-\alpha_1)}\, .
\ee
Requiring additionally that
\bea\label{vmn3}
&&\nu_2+\mu_1=B\, ,\\ \nonumber
&&\nu_2+\mu_2=D\, ,\\ \nonumber
&&\nu_1+\mu_3=E\, ,\\ \nonumber
&&\nu_3+\mu_3=F\, .
\eea
where these equalities as before understood in a sense, that odd sums identified with the NS sector, and even sums
identified with the Ramond sectors,
we get
\be\label{formbas}
F^{\cal I}_{0,\alpha_t}\left[\begin{array}{cc}
\alpha_3& \alpha_1\\
\alpha_3& \alpha_1 \end{array}\right]=C_{\cal I}(\alpha_t,\alpha_1,\alpha_3){W_{NS}(Q)W_L(\alpha_t)
\over \pi W_E(Q-\alpha_1)W_F(Q-\alpha_3)}\, .
\ee

\subsection{$\alpha_t\to 0$ limit}
\label{qw4}
Consider the same fusing matrix, but parametrized  in the form
\be\label{fmnr2}
F^{\cal I}_{\alpha_s,\alpha_t}\left[\begin{array}{cc}
\alpha_3& \alpha_2\\
\alpha_4& \alpha_1 \end{array}\right]=
{{\cal R}^{\cal I}\over i}\int_{-i\infty}^{i\infty}d\tau J^{\cal I}_{\alpha_s,\alpha_t}\left[\begin{array}{cc}
\alpha_3& \alpha_2\\
\alpha_4& \alpha_1 \end{array}\right]
\ee

with
\bea
&&{\cal R}^{\cal I}=\\  \nonumber
&&{\Gamma_E(2Q-\alpha_t-\alpha_2-\alpha_3)\Gamma_{NS}(Q-\alpha_t+\alpha_3-\alpha_2)\Gamma_E(Q+\alpha_t-\alpha_2-\alpha_3)
\Gamma_{NS}(\alpha_3+\alpha_t-\alpha_2)\over
\Gamma_A(2Q-\alpha_1-\alpha_s-\alpha_2)\Gamma_B(Q-\alpha_s-\alpha_2+\alpha_1)\Gamma_C(Q-\alpha_1-\alpha_2+\alpha_s)
\Gamma_D(\alpha_s+\alpha_1-\alpha_2)}\\  \nonumber
&\times&{\Gamma_{NS}(Q-\alpha_t-\alpha_1+\alpha_4)\Gamma_F(\alpha_1+\alpha_4-\alpha_t)\Gamma_{NS}(\alpha_t+\alpha_4-\alpha_1)
\Gamma_F(\alpha_t+\alpha_1+\alpha_4-Q)\over
\Gamma_B(Q-\alpha_s-\alpha_3+\alpha_4)\Gamma_C(\alpha_3+\alpha_4-\alpha_s)\Gamma_D(\alpha_s+\alpha_4-\alpha_3)
\Gamma_A(\alpha_s+\alpha_3+\alpha_4-Q)}\\  \nonumber
&&\times {\Gamma_{\rm L}(2Q-2\alpha_s)\Gamma_{\rm L}(2\alpha_s)\over \Gamma_{\rm NS }(Q-2\alpha_t)\Gamma_{\rm NS}(2\alpha_t-Q)}\, ,
\eea
\bea\label{jcalka}
&&J^{\cal I}_{\alpha_s,\alpha_t}\left[\begin{array}{cc}
\alpha_3& \alpha_2\\
\alpha_4& \alpha_1 \end{array}\right]=\\ \nonumber
&& {S_{\nu_1}(Q+\tau-\alpha_1)S_{\rm K}(\tau+\alpha_4+\alpha_2-\alpha_3)S_{\nu_2}(\tau+\alpha_1)S_{\nu_3}(\tau+\alpha_4+\alpha_2+\alpha_3-Q)\over
S_{\mu_1+1}(Q+\tau+\alpha_4-\alpha_t)S_{K}(\tau+\alpha_4+\alpha_t)S_{\mu_2+1}(Q+\tau+\alpha_2-\alpha_s)S_{\mu_3+1}(\tau+\alpha_2+\alpha_s)}\\ \nonumber
&+&\eta{S_{\nu_1+1}(Q+\tau-\alpha_1)S_{\rm K+1}(\tau+\alpha_4+\alpha_2-\alpha_3)S_{\nu_2+1}(\tau+\alpha_1)S_{\nu_3+1}(\tau+\alpha_4+\alpha_2+\alpha_3-Q)\over
S_{\mu_1}(Q+\tau+\alpha_4-\alpha_t)S_{K+1}(\tau+\alpha_4+\alpha_t)S_{\mu_2}(Q+\tau+\alpha_2-\alpha_s)S_{\mu_3}(\tau+\alpha_2+\alpha_s)}\, ,
\eea
where $\eta=(-1)^{ (1+\sum_i (\nu_i+\mu_i))/2}$.

We change here notations for the capital Latin letters denoting different spin structures. This is done to keep parametrization for the capital Latin letters
in the formula for structure constants (\ref{dozz}).
Alternatively we could keep the same parametrization in
formula for fusing matrix  and change the notations in formula for structure constants.

Consider the limit
\be\label{alphat}
\alpha_t=\epsilon\to 0,\hspace{1cm} \alpha_3=\alpha_2,\hspace{1cm}
\alpha_4=\alpha_1\, .
\ee
In this limit using formulas in appendix and (\ref{dozz}) we have for the factor in front of integral

\bea
&&{\cal R}^{\cal I}\to{2\over \pi\epsilon^2C_{\cal I}(\alpha_s,\alpha_2,\alpha_1)}
{W_{NS}(0)W_E(Q-\alpha_2)W_L(Q-\alpha_s)\over W_F(\alpha_1)}
\times \\ \nonumber
&&
{S_{\rm F}(2\alpha_1)\over
S_B(Q-\alpha_s-\alpha_2+\alpha_1)S_D(\alpha_s+\alpha_1-\alpha_2)
S_{\rm E}(2\alpha_2)S_{\rm NS}(\epsilon)}\, .
\eea

Consider now the limit of the integrand (\ref{jcalka}).

In the limit (\ref{alphat}) the arguments of $S_K$'s in numerator and denominator coincide and they get canceled.

For the rest of $S$'s in this  limit we get for  $a_i$ in the argument of $S_{\nu_i}(\tau+a_i)$ and $b_i$ in the argument of $S_{\mu_i+1}(-\tau+b_i)$:
\bea\label{qabt}
&&a_1=Q-\alpha_1\, ,\hspace{2.1cm} b_1=-\alpha_1\, ,\\ \nonumber
&&a_2=\alpha_1\, ,\hspace{2.9cm} b_2=\alpha_s-\alpha_2\, ,\\ \nonumber
&&a_3=2\alpha_2+\alpha_1-Q\, ,\hspace{1cm} b_3=Q-\alpha_2-\alpha_s\, .\\ \nonumber
\eea
From (\ref{qabt}) we easily obtain:
\bea\label{ta1b1}
a_1+b_1=Q-2\alpha_1\, ,\\ \nonumber
a_1+b_2=Q-\alpha_1+\alpha_s-\alpha_2\, ,\\ \nonumber
a_1+b_3=2Q-\alpha_1-\alpha_s-\alpha_2\, ,
\eea
\bea\label{ta2b2}
a_2+b_1=\epsilon\, ,\\ \nonumber
a_2+b_2=\alpha_1+\alpha_s-\alpha_2\, ,\\ \nonumber
a_2+b_3=Q-\alpha_2-\alpha_s+\alpha_1\, ,
\eea
\bea\label{ta3b3}
a_3+b_1=2\alpha_2-Q\, ,\\ \nonumber
a_3+b_2=\alpha_2+\alpha_1+\alpha_s-Q\, ,\\ \nonumber
a_3+b_3=\alpha_2+\alpha_1-\alpha_s\, .
\eea
Note that
\bea
a_1+b_3=Q-(a_3+b_2)\, ,\\ \nonumber
a_1+b_2=Q-(a_3+b_3)\, ,
\eea
and
\be\label{balqq}
\sum_i(a_i+b_i)=Q\, .
\ee
Assume that
\bea\label{tmn1}
&&\nu_1+\mu_3=\nu_3+\mu_2\quad \rm{mod}\; 2\, ,\\ \nonumber
&&\nu_1+\mu_2=\nu_3+\mu_3\quad \rm{mod}\; 2\, ,\\ \nonumber
&&\nu_2+\mu_1=1\quad \rm{mod}\; 2\, .
\eea
Under these conditions we get from the theorem (\ref{sinteg}) ,
using formulas (\ref{ta1b1})-(\ref{tmn1})
\be
{1\over i}\int_{-i\infty}^{i\infty}d\tau J^{\cal I}_{\alpha_s,\alpha_t}\left[\begin{array}{cc}
\alpha_3& \alpha_2\\
\alpha_4& \alpha_1 \end{array}\right]={2S_{\nu_2+\mu_2}(\alpha_1+\alpha_s-\alpha_2)
S_{\nu_3+\mu_1}(2\alpha_2-Q)S_{\rm NS}(\epsilon)
\over S_{\nu_1+\mu_1}(2\alpha_1)S_{\nu_2+\mu_3}(\alpha_2+\alpha_s-\alpha_1)}\, .
\ee
Requiring additionally that

\bea\label{tmn2}
&&\nu_2+\mu_3=B\, ,\\ \nonumber
&&\nu_2+\mu_2=D\, ,\\ \nonumber
&&\nu_3+\mu_1=E\, ,\\ \nonumber
&&\nu_1+\mu_1=F\, ,
\eea
where these equalities as before understood in a sense, that odd sums identified with the NS sector, and even sums
identified with the Ramond sectors,
we get
\bea\label{fastf}
\tilde{F}^{\cal I}_{\alpha_s,\epsilon}\left[\begin{array}{cc}
\alpha_2& \alpha_2\\
\alpha_1& \alpha_1 \end{array}\right]={\rm lim}_{\epsilon\to 0}\epsilon^2 F^{\cal I}_{\alpha_s,\epsilon}\left[\begin{array}{cc}
\alpha_2& \alpha_2\\
\alpha_1& \alpha_1 \end{array}\right]={4\over \pi C_{\cal I}(\alpha_s,\alpha_2,\alpha_1)}
{W_{NS}(0)W_L(Q-\alpha_s)\over W_F(\alpha_1)W_E(\alpha_2)}\, .\quad
\eea

\section{NS sector fusion matrix}
\label{qw5}
Recall that structure constants in the NS sector are given by eq. (\ref{dozz1}) and (\ref{dozz2})
and fusion matrix by (\ref{fusja}).

Remember that $NS=1,\, {\rm mod}\, 2$ and $R=0,\, {\rm mod}\, 2$.
Putting $A=B=C=D=L=E=F=NS$, $\nu_1=\nu_2=\nu_3=1$, $\mu_1=\mu_2=\mu_3=0$, and using (\ref{formbas}),
we obtain for the $(i=1,j=1)$ component of the NS sector fusing matrices in the limit (\ref{lim-s})
\be\label{falpha}
F_{0,\alpha_t}\left[\begin{array}{cc}
\alpha_3& \alpha_1\\
\alpha_3& \alpha_1 \end{array}\right]^1_1=C_{NS}(\alpha_t,\alpha_1,\alpha_3){W_{NS}(Q)W_{NS}(\alpha_t)
\over \pi W_{NS}(Q-\alpha_1)W_{NS}(Q-\alpha_3)}\, .
\ee
Putting $A=B=C=D=R$, $L=E=F=NS$, $\nu_1=\nu_2=\nu_3=1$, $\mu_1=\mu_2=1$, $\mu_3=0$, and using (\ref{formbas}), we obtain
for the $(i=2,j=1)$ component of the NS sector fusing matrices in the limit (\ref{lim-s})

\be
F_{0,\alpha_t}\left[\begin{array}{cc}
\alpha_3& \alpha_1\\
\alpha_3& \alpha_1 \end{array}\right]^2_1=\tilde{C}_{NS}(\alpha_t,\alpha_1,\alpha_3){W_{NS}(Q)W_{NS}(\alpha_t)
\over \pi W_{NS}(Q-\alpha_1)W_{NS}(Q-\alpha_3)}\, .
\ee
It is obvious to see that both choices of the $\nu_i$ and $\mu_i$
satisfy the conditions (\ref{vmn1}), (\ref{vmn2}), (\ref{vmn3}).

Putting $A=B=C=D=L=E=F=NS$, $\nu_1=\nu_2=\nu_3=1$, $\mu_1=\mu_2=\mu_3=0$, and using (\ref{fastf}),
we obtain for the $(i=1,j=1)$ component of the NS fusing matrices in the limit (\ref{alphat})
\bea
\tilde{F}_{\alpha_s,0}\left[\begin{array}{cc}
\alpha_2& \alpha_2\\
\alpha_1& \alpha_1 \end{array}\right]^1_1={\rm lim}_{\epsilon\to 0}\epsilon^2 F_{\alpha_s,\epsilon}\left[\begin{array}{cc}
\alpha_2& \alpha_2\\
\alpha_1& \alpha_1 \end{array}\right]^1_1={4\over \pi C_{NS}(\alpha_s,\alpha_2,\alpha_1)}
{W_{NS}(0)W_{NS}(Q-\alpha_s)\over W_{NS}(\alpha_1)W_{NS}(\alpha_2)}\, .\quad
\eea
Putting $A=B=C=D=R$, $L=E=F=NS$, $\nu_1=\nu_2=\nu_3=1$, $\mu_1=0$, $\mu_2=\mu_3=1$, and using (\ref{fastf}),
we obtain for the $(i=1,j=2)$ component of the NS fusing matrix in the limit (\ref{alphat})
\bea
\tilde{F}_{\alpha_s,0}\left[\begin{array}{cc}
\alpha_2& \alpha_2\\
\alpha_1& \alpha_1 \end{array}\right]^1_2={\rm lim}_{\epsilon\to 0}\epsilon^2F_{\alpha_s,\epsilon}\left[\begin{array}{cc}
\alpha_2& \alpha_2\\
\alpha_1& \alpha_1 \end{array}\right]^1_2={4\over \pi \tilde{C}_{NS}(\alpha_s,\alpha_2,\alpha_1)}
{W_{NS}(0)W_{NS}(Q-\alpha_s)\over W_{NS}(\alpha_1)W_{NS}(\alpha_2)}\, .\quad
\eea
It is again obvious to see that both sets of the values of $\nu_i$ and $\mu_i$ satisfy the conditions (\ref{vmn2}), (\ref{tmn1}) and
(\ref{tmn2}).

Note also the relations:
\be\label{mose1}
F_{0,\alpha_s}\left[\begin{array}{cc}
\alpha_1& \alpha_2\\
\alpha_1& \alpha_2 \end{array}\right]^1_1
\tilde{F}_{\alpha_s,0}\left[\begin{array}{cc}
\alpha_2& \alpha_2\\
\alpha_1& \alpha_1 \end{array}\right]^1_1={S(0)S(\alpha_s)
\over \pi^2S(\alpha_1)S(\alpha_2)}\, ,
\ee
\be\label{mose2}
F_{0,\alpha_s}\left[\begin{array}{cc}
\alpha_1& \alpha_2\\
\alpha_1& \alpha_2 \end{array}\right]^2_1
\tilde{F}_{\alpha_s,0}\left[\begin{array}{cc}
\alpha_2& \alpha_2\\
\alpha_1& \alpha_1 \end{array}\right]^1_2={S(0)S(\alpha_s)
\over \pi^2S(\alpha_1)S(\alpha_2)}\, ,
\ee
where $S(\alpha)=\sin\pi b(\alpha-Q/2)\sin\pi {1\over b}(\alpha-Q/2)$.

Remembering the relation (\ref{motrdi1}) and that the vacuum field is given by the pair $(1,1)$ we see that the function $S(\alpha)$
coincide with the vacuum component of the matrix of modular transformations. 
We see that the relations  (\ref{falpha})-(\ref{mose2}) indeed have the structure of the equations (\ref{foifio}),(\ref{cpijnx}) and
(\ref{cpij}).

\section{Fusion matrix in the Ramond sector}
\label{qw6}
The fusion matrix in the Ramond sector unfortunately is not known in general.
Although for some attempts see   \cite{Chorazkiewicz:2011zd}.
But for the degenerate primaries (\ref{degnm}) fusion matrix can be computed
via direct solutions of the corresponding differential equation for conformal blocks.
In particular the necessary elements of the fusion matrix when one of the entries is the simplest
degenerate field $R_{-b/2}$ are computed in \cite{Ahn:2002ev,Fukuda:2002bv}.
The degenerate field $R_{-b/2}$ possesses the OPE:
\be\label{nc1n}
N_{\alpha}R_{-b/2}=C^{R_{\alpha-b/2}}_{N_{\alpha}R_{-b/2}}R_{\alpha-b/2}+
C^{R_{\alpha+b/2}}_{N_{\alpha}R_{-b/2}}R_{\alpha+b/2}\, ,
\ee
\be\label{nc2n}
R_{\alpha}R_{-b/2}=C^{N_{\alpha-b/2}}_{R_{\alpha}R_{-b/2}}
N_{\alpha-b/2}+C^{N_{\alpha+b/2}}_{R_{\alpha}R_{-b/2}}N_{\alpha+b/2}\, .
\ee
The corresponding structure constant can be computed in the Coulomb gas formalism using the screening integrals:
\be\label{st1}
C^{R_{\alpha-b/2}}_{N_{\alpha}R_{-b/2}}=1\, ,
\ee
\be\label{st2}
C^{R_{\alpha+b/2}}_{N_{\alpha}R_{-b/2}}=\pi\mu b^2\gamma(bQ/2)\gamma(1-b\alpha)\gamma(b\alpha-bQ/2)=
{{\cal G}_{NS}(\alpha)\over {\cal G}_{R}(\alpha+b/2)}\, ,
\ee
\be\label{st3}
C^{N_{\alpha-b/2}}_{R_{\alpha}R_{-b/2}}=1\, ,
\ee
\be\label{st4}
C^{N_{\alpha+b/2}}_{R_{\alpha}R_{-b/2}}=2i \pi\mu b^2\gamma(bQ/2)\gamma(1/2-b\alpha)\gamma(b\alpha-b^2/2)=
2i{{\cal G}_{R}(\alpha)\over {\cal G}_{NS}(\alpha+b/2)}\, .
\ee
The fusion matrices can be computed having explicit expression of the conformal blocks with degenerate entries:

\be\label{f1}
F_{R_{\alpha-b/2},0}\left[\begin{array}{cc}
R_{-b/2}& R_{-b/2}\\
N_{\alpha}& N_{\alpha} \end{array}\right]={\Gamma(\alpha b -b^2/2+1/2)\Gamma(-b^2)\over \Gamma(\alpha b-b^2)\Gamma(1/2-b^2/2)}\, ,
\ee
\be\label{f2}
F_{R_{\alpha+b/2},0}\left[\begin{array}{cc}
R_{-b/2}& R_{-b/2}\\
N_{\alpha}& N_{\alpha} \end{array}\right]={\Gamma(-\alpha b +b^2/2+3/2)\Gamma(-b^2)\over \Gamma(1-\alpha b)\Gamma(1/2-b^2/2)}\, ,
\ee
\be\label{f3}
F_{N_{\alpha-b/2},0}\left[\begin{array}{cc}
R_{-b/2}& R_{-b/2}\\
R_{\alpha}& R_{\alpha} \end{array}\right]={\Gamma(\alpha b -b^2/2)\Gamma(-b^2)\over \Gamma(\alpha b-b^2-1/2)\Gamma(1/2-b^2/2)}\, ,
\ee
\be\label{f4}
F_{N_{\alpha+b/2},0}\left[\begin{array}{cc}
R_{-b/2}& R_{-b/2}\\
R_{\alpha}& R_{\alpha} \end{array}\right]={\Gamma(-\alpha b +b^2/2+1)\Gamma(-b^2)\over 2i \Gamma(1/2-\alpha b)\Gamma(1/2-b^2/2)}\, .
\ee
It is an easy exercise to check that the values of the structure constants (\ref{st1})-(\ref{st4})
and fusion matrices (\ref{f1})-(\ref{f4}) satisfy the relations:
\be\label{cf1}
C^{R_{\alpha-b/2}}_{N_{\alpha}R_{-b/2}}F_{R_{\alpha-b/2},0}\left[\begin{array}{cc}
R_{-b/2}& R_{-b/2}\\
N_{\alpha}& N_{\alpha} \end{array}\right]={\Gamma(\alpha b -b^2/2+1/2)\Gamma(-b^2)\over \Gamma(\alpha b-b^2)\Gamma(1/2-b^2/2)}=
{W_{NS}(0)W_R(\alpha-b/2)\over W_{NS}(\alpha)W_R(-b/2)}\, ,
\ee

\be\label{cf2}
C^{R_{\alpha+b/2}}_{N_{\alpha}R_{-b/2}}F_{R_{\alpha+b/2},0}\left[\begin{array}{cc}
R_{-b/2}& R_{-b/2}\\
N_{\alpha}& N_{\alpha} \end{array}\right]={\pi\mu b^2\gamma(bQ/2)\Gamma(-b^2)\Gamma(\alpha b -b^2/2-1/2)\over
\Gamma(1/2-b^2/2)\Gamma(\alpha b)}={W_{NS}(0)W_R(\alpha+b/2)\over W_{NS}(\alpha)W_R(-b/2)}\, ,
\ee

\be\label{cf3}
C^{N_{\alpha-b/2}}_{R_{\alpha}R_{-b/2}}F_{N_{\alpha-b/2},0}\left[\begin{array}{cc}
R_{-b/2}& R_{-b/2}\\
R_{\alpha}& R_{\alpha} \end{array}\right]={\Gamma(\alpha b -b^2/2)\Gamma(-b^2)\over \Gamma(\alpha b-b^2-1/2)\Gamma(1/2-b^2/2)}=
{W_{NS}(0)W_{NS}(\alpha-b/2)\over W_R(\alpha)W_R(-b/2)}\, ,
\ee
\be\label{cf4}
C^{N_{\alpha+b/2}}_{R_{\alpha}R_{-b/2}}F_{N_{\alpha+b/2},0}\left[\begin{array}{cc}
R_{-b/2}& R_{-b/2}\\
R_{\alpha}& R_{\alpha} \end{array}\right]={\pi\mu b^2\gamma(bQ/2)\Gamma(\alpha b -b^2/2)\Gamma(-b^2)\over \Gamma(\alpha b+1/2)\Gamma(1/2-b^2/2)}=
{W_{NS}(0)W_{NS}(\alpha+b/2)\over W_R(\alpha)W_R(-b/2)}\, .
\ee
One expects that similar relations should hold also for general expressions of the corresponding elements of fusion matrix in the RR sector.
 For example  the fusions matrix with four RR entries
should satisfy the relations
\bea\label{fastf3}
{\rm lim}_{\epsilon\to 0}\epsilon^2F_{N_{\alpha_s},N_{\epsilon}}\left[\begin{array}{cc}
R_{\alpha_2}& R_{\alpha_2}\\
R_{\alpha_1}& R_{\alpha_1} \end{array}\right]={4\over \pi (C_{R}(\alpha_s|\alpha_2,\alpha_1)+\tilde{C}_{R}(\alpha_s|\alpha_1,\alpha_2))}
{W_{NS}(0)W_{NS}(Q-\alpha_s)\over W_R(\alpha_1)W_R(\alpha_2)}\, ,\quad\quad
\eea

\be\label{formbas3}
F_{0,N_{\alpha_t}}\left[\begin{array}{cc}
R_{\alpha_3}& R_{\alpha_1}\\
R_{\alpha_3}& R_{\alpha_1} \end{array}\right]=(C_{R}(\alpha_t|\alpha_1,\alpha_3)+\tilde{C}_{R}(\alpha_t|\alpha_1,\alpha_3)){W_{NS}(Q)W_{NS}(\alpha_t)
\over \pi W_R(Q-\alpha_1)W_R(Q-\alpha_3)}\, .
\ee
One can hope that constraints like (\ref{fastf3}) and (\ref{formbas3}) may help to obtain the general expressions
for the corresponding elements of the fusion matrix. 

\section{Defects in Super-Liouville theory}
\label{qw7}
Two-point functions with a defect $X$ insertion can be written as
\be\label{twopff}
\langle\Phi_{i}(z_1,\bar{z}_1)X\Phi_{i}(z_2,\bar{z}_2)\rangle={D^{i}\over
(z_1-z_2)^{2\Delta_i}(\bar{z}_1-\bar{z}_2)^{2\Delta_i}}\, ,
\ee

where
\be\label{conjik}
D^{i}={\cal D}^{i}C_{ii}
\ee
and $C_{ii}$ is a two-point function.
They satisfy the Cardy-Lewellen equation for defects \cite{Sarkissian:2009aa,Sarkissian:2011tr,Petkova:2001ag,Poghosyan:2015oua}
\bea\label{defalg}
\sum_k D^0D^{k}\left(C_{ij}^{k}
 F_{k0}\left[\begin{array}{cc}
j&j\\
i&i\end{array}\right]\right)^2=
D^{i}D^{j}\, .\hspace{1cm}
\eea



Denote
\be
D_{NS}(\alpha)=\langle N_{\alpha} X N_{\alpha}\rangle\, ,
\ee
\be
D_{R}(\alpha)=\langle R_{\alpha} X R_{\alpha}\rangle\, .
\ee
Let us take $j=R_{-b/2}$. Using (\ref{nc1n}), (\ref{nc2n}) and (\ref{cf1})-(\ref{cf4}) one can obtain:
\be\label{psik}
\Psi_{NS}(\alpha)\Psi_R(-b/2)=\Psi_R(\alpha-b/2)+\Psi_R(\alpha+b/2)\, ,
\ee
\be\label{psik2}
\Psi_{R}(\alpha)\Psi_R(-b/2)=\Psi_{NS}(\alpha-b/2)+\Psi_{NS}(\alpha+b/2)\, ,
\ee
where
\be\label{dkdol}
{D_{NS}(\alpha)\over D_{NS}(0)}=\Psi_{NS}(\alpha)\left({W_{NS}(0)\over W_{NS}(\alpha)}\right)^2\, ,
\ee
\be\label{dkdoll}
{D_R(\alpha)\over D_{NS}(0)}=\Psi_R(\alpha)\left({W_{NS}(0)\over W_R(\alpha)}\right)^2\, .
\ee
The solution of the equations (\ref{psik}) and (\ref{psik2}) is
\be\label{psb1}
\Psi_{NS}(\alpha; m,n)={\sin(\pi m b^{-1}(\alpha- Q/2))\sin(\pi nb (\alpha-Q/2))\over \sin(\pi {mb^{-1} Q\over 2})\sin({\pi nb Q\over 2})}\, ,
\ee

\be\label{psb2}
\Psi_{R}(\alpha; m,n)={\sin(\pi m({1\over 2}+ b^{-1}(\alpha- Q/2)))\sin(\pi n({1\over 2}+b (\alpha-Q/2)))\over \sin({\pi mb^{-1} Q\over 2})\sin({\pi nb Q\over 2})}\, ,
\ee
with $m-n$ is even.

Substituting (\ref{psb1}) and (\ref{psb2}) in  (\ref{dkdol}) and (\ref{dkdoll}) we obtain
\be
D_{NS}(\alpha;m,n)={\sin(\pi m b^{-1}(\alpha- Q/2))\sin(\pi nb (\alpha-Q/2))\over W_{NS}(\alpha)^2}\, ,
\ee
\be
D_{R}(\alpha;m,n)={\sin(\pi m({1\over 2}+ b^{-1}(\alpha- Q/2)))\sin(\pi n({1\over 2}+b (\alpha-Q/2)))\over W_R(\alpha)^2}\, .
\ee
Dividing by two-point functions (\ref{dna}) and (\ref{dnr}) we obtain
\be
{\cal D}_{NS}(\alpha;m,n)={\sin(\pi m b^{-1}(\alpha- Q/2))\sin(\pi nb (\alpha-Q/2))\over \sin(\pi  b^{-1}(\alpha- Q/2))\sin(\pi b (\alpha-Q/2))}\, ,
\ee
\be
{\cal D}_{R}(\alpha;m,n)={\sin(\pi m({1\over 2}+ b^{-1}(\alpha- Q/2)))\sin(\pi n({1\over 2}+b (\alpha-Q/2)))\over \cos(\pi  b^{-1}(\alpha- Q/2))\cos(\pi b (\alpha-Q/2))}\, .
\ee
To obtain the continuous family of defects we use the strategy developed in \cite{Fateev:2000ik,Fateev:2010za}.
Namely consider $D_{R}(-b/2)$ as a parameter characterizing a defect.
More precisely we define
\be
A={D_{R}(-b/2)\over D_{NS}(0)}\left({W_R(-b/2)\over W_{NS}(0)}\right)^2\, .
\ee
Denoting also
\be\label{dkdol1}
D_{NS}(\alpha)={\tilde{\Psi}_{NS}(\alpha)\over W_{NS}(\alpha)^2}\, ,
\ee
\be\label{dkdoll2}
D_R(\alpha)={\tilde{\Psi}_R(\alpha)\over W_R(\alpha)^2}\, .
\ee
we obtain
\be\label{psikk}
A\tilde{\Psi}_{NS}(\alpha)=\tilde{\Psi}_R(\alpha-b/2)+\tilde{\Psi}_R(\alpha+b/2)\, ,
\ee
\be\label{psikk2}
A\tilde{\Psi}_{R}(\alpha)=\tilde{\Psi}_{NS}(\alpha-b/2)+\tilde{\Psi}_{NS}(\alpha+b/2)\, ,
\ee

The solution of (\ref{psikk}) and (\ref{psikk2}) is given by
\be\label{psoir}
\tilde{\Psi}_{NS}(\alpha;u)=\cosh(\pi(2\alpha-Q)u)\, ,
\ee
\be\label{psoim}
\tilde{\Psi}_{R}(\alpha;u)=\cosh(\pi(2\alpha-Q)u)\, ,
\ee
with a parameter $u$ related to $A$ by
\be\label{rela}
2\cosh 2\pi b u=A\, .
\ee
Substituting (\ref{psoir}) and (\ref{psoim}) in  (\ref{dkdol1}) and (\ref{dkdoll2}) we obtain
\be
D_{NS}(\alpha;u)={\cosh(\pi(2\alpha-Q)u)\over W_{NS}(\alpha)^2}\, ,
\ee
\be
D_{R}(\alpha;u)={\cosh(\pi(2\alpha-Q)u)\over W_R(\alpha)^2}\, .
\ee
Dividing by two-point functions (\ref{dna}) and (\ref{dnr}) we obtain
\be
{\cal D}_{NS}(\alpha;u)={\cosh(\pi(2\alpha-Q)u)\over \sin(\pi  b^{-1}(\alpha- Q/2))\sin(\pi b (\alpha-Q/2))}\, ,
\ee
\be
{\cal D}_{R}(\alpha;u)={\cosh(\pi(2\alpha-Q)u)\over \cos(\pi  b^{-1}(\alpha- Q/2))\cos(\pi b (\alpha-Q/2))}\, .
\ee

\section{Discussion}
\label{qwd}

The methods of this paper can be useful to construct fusion matrix in the parafermionic Liouville field theory \cite{Bershtein:2010wz}.
Parafermionic Liouville field theory is the simplest generalization of the supersymmetric Liouville theory.
Whereas the supersymmetric Liouville theory is the Liouville field theory coupled to the Ising model,
the parafermionic Liouville field theory is the Liouville field theory coupled to the parafermions.
The structure constants of the parafermionic Liouville field theory at the level $N$ can be written using the following generalization of
the $\Upsilon_{NS}$ and $\Upsilon_{R}$ functions 
\be
\Upsilon^{(N)}_k(x)=\prod_{j=1}^{N-k}\Upsilon_b\left({x+kb^{-1}+(j-1)Q\over N}\right)
\prod_{j=N-k+1}^{N}\Upsilon_b\left({x+(k-N)b^{-1}+(j-1)Q\over N}\right)\, .
\ee

It is easy to check that these functions can be written as
\be
\Upsilon^{(N)}_k(x)={1\over \Gamma^{(N)}_k(x)\Gamma^{(N)}_{N-k}(Q-x)}\, ,
\ee
where
\be
\Gamma^{(N)}_k(x)=\prod_{j=1}^{N-k}\Gamma_b\left({x+kb^{-1}+(j-1)Q\over N}\right)
\prod_{j=N-k+1}^{N}\Gamma_b\left({x+(k-N)b^{-1}+(j-1)Q\over N}\right)\, .
\ee
The functions $\Gamma^{(N)}_k(x)$ have the property
\be
{\Gamma^{(N)}_k(x+Q)\over \Gamma^{(N)}_k(x)}=W_k(x)={2\pi b^{(b-b^{-1})x\over N}b^{{2k\over N}-1}\over \Gamma\left({k\over N}+{bx\over N}\right)
\Gamma\left(1-{k\over N}+{b^{-1}x\over N}\right)}\, ,
\ee
which is very similar to (\ref{gns}) and (\ref{grr}). Recall that these properties played crucial role in calculations in section
\ref{qw2}.
Therefore one can try to write
fusion matrix in the paraferminonic Liouville field theory using the corresponding para version
of the double Gamma and double Sine functions and matching the relations (\ref{foifio}), (\ref{cpijnx}), (\ref{cpij})
with the parafermionic Liouville structure constants found in \cite{Bershtein:2010wz}.

It is well known that in the AGT correspondence Wilson lines in the $N=2$ $SU(N)$ $(SU(2))$ superconformal gauge theory on $S^4$  correspond to
topological defects in Toda (Liouville) conformal field theory \cite{Drukker:2010jp}.
On the other hand it is found that $N=2$ $SU(N)$ $(SU(2))$ superconformal gauge theory on $S^4/\mathbb{Z}_p$
correspond to parafermionic Toda (Liouville) field theories.
In particular  ${N=2}$ $SU(2)$ superconformal gauge theory on $S^4/\mathbb{Z}_2$ correspond to supersymmetric Liouville field
theory. Thus having
the topological defects in superLiouville theory one can test the AGT correspondence
of $SU(2)$ superconformal gauge theory on $S^4/\mathbb{Z}_2$ with supersymmetric Liouville field theory
in the presence of the Wilson lines.

The Lagrangian of the $N=1$ super Liouville field theory with the topological defect was introduced in 
\cite{Aguirre:2013zfa}. In \cite{Poghosyan:2015oua} the light and heavy semiclassical limits were used to match
two-point correlation function with the Lagrangian approach for the bosonic Liouville theory in the presence of the defects.
It is an interesting task to match, using various semiclassical techniques, the results of section \ref{qw7} with
the Lagrangian of \cite{Aguirre:2013zfa}.

\section*{Acknowledgments}
The work of G.S. was partially supported by the Armenian SCS  grant 15T-1C308
and by ICTP within  NET68 and OEA-AC-100 projects.
The work of H.P. and G.S. was supported also by the ANSEF grant hepth-4208.
\\
\vspace*{3pt}

\appendix
\section{Useful formulae}
{\bf The function} $\Gamma_b(x)$

The function $\Gamma_b(x)$ is a close relative of the double Gamma function studied in \cite{gam1,gam2}.
It can be defined by means of the integral representation
\be
\log \Gamma_b(x)=\int_0^{\infty}{dt\over t}\left({e^{-xt}-e^{-Qt/2}\over (1-e^{-bt})(1-e^{-t/b})}-
{(Q-2x)^2\over 8e^t}-{Q-2x\over t}\right)\, .
\ee

Important properties of $\Gamma_b(x)$ are
\begin{enumerate}
\item
Functional equation: $\Gamma_b(x+b)=\sqrt{2\pi}b^{bx-{1\over 2}}\Gamma^{-1}(bx)\Gamma_b(x)$.
\item
Analyticity:  $\Gamma_b(x)$ is meromorphic, poles: $x=-nb-mb^{-1}, n,m\in \mathbb{Z}^{\geq 0}$.
\item
Self-duality:  $\Gamma_b(x)= \Gamma_{1/b}(x)$.
\end{enumerate}

The function $\Upsilon_b(x)$ may be defined in terms of $\Gamma_b(x)$ as follows
\be
\Upsilon_b(x)={1\over  \Gamma_b(x) \Gamma_b(Q-x)}\, .
\ee
It has the following property:
\be
\Upsilon_b'(0)=\Upsilon_b(b)={2\pi \over \Gamma_b^2(Q)}\, .
\ee
In the super Liouville theory are important the functions
\be
\Gamma_{1}(x)\equiv\Gamma_{\rm NS}(x)=\Gamma_b\left({x\over 2}\right)\Gamma_b\left({x+Q\over 2}\right)\, ,
\ee

\be
\Gamma_{0}(x)\equiv\Gamma_{\rm R}(x)=\Gamma_b\left({x+b\over 2}\right)\Gamma_b\left({x+b^{-1}\over 2}\right)\, .
\ee
They have the  properties:
\be\label{gns}
{\Gamma_{\rm NS}(2\alpha)\over \Gamma_{\rm NS}(2\alpha-Q)}=W_{NS}(\alpha)\lambda^{{Q-2\alpha\over 2b}}\, ,
\ee

\be\label{grr}
{\Gamma_{\rm R}(2\alpha)\over \Gamma_{\rm R}(2\alpha-Q)}=W_{\rm R}(\alpha)\lambda^{{Q-2\alpha\over 2b}}\, ,
\ee
where $W_{NS}(\alpha)$, $W_{\rm R}(\alpha)$ are defined in (\ref{wn}) and (\ref{wr}), and $\lambda=\pi\mu\gamma\left({bQ\over 2}\right)b^{1-b^2}$.

$\Gamma_{\rm NS}(x)$ has a pole in zero:
\be
\Gamma_{\rm NS}(x)\sim {\Gamma_{\rm NS}(Q)\over \pi x}\, .
\ee
The structure constants in the super Liouville theory are defined in terms of the functions:
\be
\Upsilon_{1}(x)\equiv\Upsilon_{\rm NS}(x)=\Upsilon_{b}\left({x\over 2}\right)\Upsilon_{b}\left({x+Q\over 2}\right)={1\over \Gamma_{\rm NS}(x) \Gamma_{\rm NS}(Q-x)}\, ,
\ee

\be
\Upsilon_{0}(x)\equiv\Upsilon_{\rm R}(x)=\Upsilon_{b}\left({x+b\over 2}\right)\Upsilon_{b}\left({x+b^{-1}\over 2}\right)={1\over \Gamma_{\rm R}(x) \Gamma_{\rm R}(Q-x)}\, .
\ee

They have the  properties:

\be
{\Upsilon_{\rm NS}(2x)\over \Upsilon_{\rm NS}(2x-Q)}={\cal G}_{NS}(x)\lambda^{-{Q-2x\over b}}\, ,
\ee

\be
{\Upsilon_{\rm R}(2x)\over \Upsilon_{\rm R}(2x-Q)}={\cal G}_R(x)\lambda^{-{Q-2x\over b}}\, ,
\ee
where ${\cal G}_{NS}(x)$ and ${\cal G}_R(x)$ are defined in (\ref{dna}) and (\ref{dnr}).

The zeroes of $\Upsilon_{\rm NS}$, $\Upsilon_{\rm R}$ are
\be
\Upsilon_{\rm NS}(x)=0\quad {\rm at} \quad x=-mb-nb^{-1},\quad x=Q+mb+nb^{-1}\quad (m+n\quad {\rm even})\, ,
\ee
\be
\Upsilon_{\rm R}(x)=0\quad {\rm at} \quad x=-mb-nb^{-1},\quad x=Q+mb+nb^{-1}\quad (m+n\quad {\rm odd})\, .
\ee
We need also the values of the derivative $\Upsilon_{\rm NS}'(0)$ in zero:
\be
\Upsilon_{\rm NS}'(0)={\pi \over \Gamma_{\rm NS}^2(Q)}\, .
\ee
To write fusion matrix we need also the functions:
\be
S_{1}(x)\equiv S_{\rm NS}(x)={\Gamma_{\rm NS}(x) \over \Gamma_{\rm NS}(Q-x)}\, ,
\ee

\be
S_{0}(x)\equiv S_{\rm R}(x)={\Gamma_{\rm R}(x) \over \Gamma_{\rm R}(Q-x)}\, .
\ee
They have the  properties:
\be
{S_{\rm NS}(2x)\over S_{\rm NS}(2x-Q)}=W_{NS}(x)W_{NS}(Q-x)\, ,
\ee

\be
{S_{\rm R}(2x)\over S_{\rm R}(2x-Q)}=W_R(x)W_R(Q-x)\, .
\ee
And finally we need the following properties which can be easily obtained from the definitions and properties above:
\be
\Gamma_{\rm A}(2Q-2\alpha)\Gamma_{\rm A}(Q-2\alpha)={W_A(Q-\alpha)\lambda^{-{Q-2\alpha\over 2b}}\over
\Upsilon_{\rm A}(2\alpha)S_{\rm A}(2\alpha)}\, ,
\ee

\be
\Gamma_{\rm A}(2\alpha-Q)\Gamma_{\rm A}(Q-2\alpha)={\lambda^{-{Q-2\alpha\over 2b}}\over
\Upsilon_{\rm A}(2\alpha)W_A(\alpha)}\, ,
\ee

\be
\Gamma_{\rm A}(2\alpha)\Gamma_{\rm A}(2\alpha-Q)={S_{\rm A}(2\alpha)\lambda^{-{Q-2\alpha\over 2b}}\over
\Upsilon_{\rm A}(2\alpha)W_A(\alpha)}\, ,
\ee

\be
\Gamma_{\rm A}(2Q-2\alpha)\Gamma_{\rm A}(2\alpha)={W_A(Q-\alpha)\lambda^{-{Q-2\alpha\over 2b}}\over
\Upsilon_{\rm A}(2\alpha)}\, ,
\ee
where $A$ takes values $NS$ or $R$.


\begin{thebibliography}{99}
\bibitem{Verlinde:1988sn}
  E.~P.~Verlinde,
  ``Fusion Rules and Modular Transformations in 2D Conformal Field Theory,''
  Nucl.\ Phys.\ B {\bf 300} (1988) 360.
\bibitem{Moore:1988ss}
  G.~W.~Moore and N.~Seiberg,
  ``Naturality in Conformal Field Theory,''
  Nucl.\ Phys.\  B {\bf 313} (1989) 16.
\bibitem{Moore:1989vd}
  G.~W.~Moore and N.~Seiberg,
  ``Lectures on RCFT,''  Published in Trieste Superstrings 1989:1-129.  Also in Banff NATO ASI 1989:263-362.
\bibitem{Moore:1988qv}
  G.~W.~Moore and N.~Seiberg,
  ``Classical and Quantum Conformal Field Theory,''
  Commun.\ Math.\ Phys.\  {\bf 123} (1989) 177.
\bibitem{Behrend:1999bn}
  R.~E.~Behrend, P.~A.~Pearce, V.~B.~Petkova and J.~B.~Zuber,
  ``Boundary conditions in rational conformal field theories,''
  Nucl.\ Phys.\  B {\bf 570} (2000) 525
  [Nucl.\ Phys.\  B {\bf 579} (2000) 707]
  \href{http://arxiv.org/abs/hep-th/9908036}{hep-th/9908036}.
\bibitem{Petkova:2000ip}
  V.~B.~Petkova and J.~B.~Zuber,
  ``Generalized twisted partition functions,''
  Phys.\ Lett.\  B {\bf 504} (2001) 157
 \href{http://arxiv.org/abs/hep-th/0011021}{hep-th/0011021}.
\bibitem{Petkova:2001ag}
  V.~B.~Petkova and J.~B.~Zuber,
  ``The Many faces of Ocneanu cells,''  Nucl.\ Phys.\ B {\bf 603} (2001) 449
  \href{http://arxiv.org/abs/hep-th/0101151}{hep-th/0101151}.  
  \bibitem{Dorn:1994xn}
  H.~Dorn and H.~J.~Otto,
  ``Two and three point functions in Liouville theory,''
  Nucl.\ Phys.\  B {\bf 429} (1994) 375
 \href{http://arxiv.org/abs/hep-th/9403141}{hep-th/9403141}.
\bibitem{Zamolodchikov:1995aa}
  A.~B.~Zamolodchikov and A.~B.~Zamolodchikov,
  ``Structure constants and conformal bootstrap in Liouville field theory,''
  Nucl.\ Phys.\  B {\bf 477} (1996) 577
   \href{http://arxiv.org/abs/hep-th/9506136}{hep-th/9506136}.
\bibitem{Ponsot:1999uf}
  B.~Ponsot and J.~Teschner,
  ``Liouville bootstrap via harmonic analysis on a noncompact quantum group,''
   \href{http://arxiv.org/abs/hep-th/9911110}{hep-th/9911110}.



\bibitem{Alday:2009aq}
  L.~F.~Alday, D.~Gaiotto and Y.~Tachikawa,
  ``Liouville Correlation Functions from Four-dimensional Gauge Theories,''
  Lett.\ Math.\ Phys.\  {\bf 91} (2010) 167
 \href{http://arxiv.org/abs/0906.3219}{arXiv:0906.3219}.
\bibitem{Belavin:2011pp}
  V.~Belavin and B.~Feigin,
  ``Super Liouville conformal blocks from N=2 SU(2) quiver gauge theories,''
  JHEP {\bf 1107} (2011) 079
  \href{http://arxiv.org/abs/1105.5800}{arXiv:1105.5800}.
\bibitem{Wyllard:2009hg}
  N.~Wyllard,
  ``$A_{N-1}$ conformal Toda field theory correlation functions from conformal N = 2 SU(N) quiver gauge theories,''
  JHEP {\bf 0911} (2009) 002
  \href{http://arxiv.org/abs/0907.2189}{arXiv:0907.2189}.
\bibitem{Nishioka:2011jk}
  T.~Nishioka and Y.~Tachikawa,
  ``Central charges of para-Liouville and Toda theories from M-5-branes,''
  Phys.\ Rev.\ D {\bf 84} (2011) 046009
  \href{http://arxiv.org/abs/1106.1172}{arXiv:1106.1172}.
\bibitem{Bershtein:2010wz}
  M.~A.~Bershtein, V.~A.~Fateev and A.~V.~Litvinov,
  ``Parafermionic polynomials, Selberg integrals and three-point correlation function in parafermionic Liouville field theory,''
  Nucl.\ Phys.\ B {\bf 847} (2011) 413
  \href{http://arxiv.org/abs/1011.4090}{arXiv:1011.4090}.
\bibitem{Bonelli:2011jx}
  G.~Bonelli, K.~Maruyoshi and A.~Tanzini,
  ``Instantons on ALE spaces and Super Liouville Conformal Field Theories,''
  JHEP {\bf 1108} (2011) 056
  \href{http://arxiv.org/abs/1106.2505}{arXiv:1106.2505}.
\bibitem{Bonelli:2011kv}
  G.~Bonelli, K.~Maruyoshi and A.~Tanzini,
  ``Gauge Theories on ALE Space and Super Liouville Correlation Functions,''
  Lett.\ Math.\ Phys.\  {\bf 101} (2012) 103
  \href{http://arxiv.org/abs/1107.4609}{arXiv:1107.4609}.
\bibitem{Bonelli:2012ny}
  G.~Bonelli, K.~Maruyoshi, A.~Tanzini and F.~Yagi,
  ``N=2 gauge theories on toric singularities, blow-up formulae and W-algebrae,''
  JHEP {\bf 1301} (2013) 014
  \href{http://arxiv.org/abs/1208.0790}{arXiv:1208.0790}.
\bibitem{Belavin:2012aa}
  A.~Belavin and B.~Mukhametzhanov,
  ``N=1 superconformal blocks with Ramond fields from AGT correspondence,''
  JHEP {\bf 1301} (2013) 178
  \href{http://arxiv.org/abs/1210.7454}{arXiv:1210.7454}.
\bibitem{Poghosian:1996dw}
  R.~H.~Poghossian,
  ``Structure constants in the N=1 superLiouville field theory,''
  Nucl.\ Phys.\ B {\bf 496} (1997) 451
  \href{http://arxiv.org/abs/hep-th/9607120}{hep-th/9607120}.
\bibitem{Rashkov:1996np}
  R.~C.~Rashkov and M.~Stanishkov,
  ``Three point correlation functions in N=1 superLiouville theory,''
  Phys.\ Lett.\ B {\bf 380} (1996) 49
  \href{http://arxiv.org/abs/hep-th/9602148}{hep-th/9602148}.
\bibitem{Hadasz:2007wi}
  L.~Hadasz,
  ``On the fusion matrix of the N=1 Neveu-Schwarz blocks,''
  JHEP {\bf 0712} (2007) 071
  \href{http://arxiv.org/abs/0707.3384}{arXiv:0707.3384}.
\bibitem{Chorazkiewicz:2008es}
  D.~Chorazkiewicz and L.~Hadasz,
  ``Braiding and fusion properties of the Neveu-Schwarz super-conformal blocks,''
  JHEP {\bf 0901} (2009) 007
  \href{http://arxiv.org/abs/0811.1226}{arXiv:0811.1226}.
\bibitem{Chorazkiewicz:2011zd}
  D.~Chorazkiewicz, L.~Hadasz and Z.~Jaskolski,
  ``Braiding properties of the N=1 super-conformal blocks (Ramond sector),''
  JHEP {\bf 1111} (2011) 060
  \href{http://arxiv.org/abs/1108.2355}{arXiv:1108.2355}.
\bibitem{Pawelkiewicz:2013wga}
  M.~Pawelkiewicz, V.~Schomerus and P.~Suchanek,
  ``The universal Racah-Wigner symbol for $U_q({\rm osp}(1|2))$,''
  JHEP {\bf 1404} (2014) 079
  \href{http://arxiv.org/abs/1307.6866}{arXiv:1307.6866}.
\bibitem{Felder:1989hq}
  G.~Felder, J.~Frohlich and G.~Keller,
  ``On the structure of unitary conformal field theory. 2. Representation
  theoretic approach,''
  Commun.\ Math.\ Phys.\  {\bf 130} (1990) 1.
\bibitem{Fuchs:2004xi}
  J.~Fuchs, I.~Runkel and C.~Schweigert,
  ``TFT construction of RCFT correlators IV: Structure constants and
  correlation functions,''
  Nucl.\ Phys.\  B {\bf 715} (2005) 539
  \href{http://arxiv.org/abs/hep-th/0412290}{hep-th/0412290}.
\bibitem{Sarkissian:2011tr}
  G.~Sarkissian,
  ``Some remarks on D-branes and defects in Liouville and Toda field theories,''  Int.\ J.\ Mod.\ Phys.\ A {\bf 27} (2012) 1250181  \href{http://arxiv.org/abs/1108.0242}{arXiv:1108.0242}.  
\bibitem{Teschner:2008qh}
  J.~Teschner,
  ``Nonrational conformal field theory,'' ``New Trends in
Mathematical Physics" (Selected contributions of the XVth ICMP),
Vladas Sidoravicius (ed.), Springer Science and Business Media B.V. 2009,
 \href{http://arxiv.org/abs/0803.0919}{arXiv:0803.0919}.
\bibitem{Vartanov:2013ima}
  J.~Teschner and G.~S.~Vartanov,
  ``Supersymmetric gauge theories, quantization of $\mathcal{M}_{\mathrm{flat}}$, and conformal field theory,''
  Adv.\ Theor.\ Math.\ Phys.\  {\bf 19} (2015) 1
  \href{http://arxiv.org/abs/1302.3778}{arXiv:1302.3778}.

\bibitem{Zamolodchikov:2001ah}
  A.~B.~Zamolodchikov and A.~B.~Zamolodchikov,
  ``Liouville field theory on a pseudosphere,''
  arXiv:hep-th/0101152.
\bibitem{Ahn:2002ev}
  C.~Ahn, C.~Rim and M.~Stanishkov,
  ``Exact one point function of N=1 superLiouville theory with boundary,''
  Nucl.\ Phys.\ B {\bf 636} (2002) 497
  \href{http://arxiv.org/abs/hep-th/0202043}{hep-th/0202043}.
\bibitem{Fukuda:2002bv}
  T.~Fukuda and K.~Hosomichi,
  ``Super Liouville theory with boundary,''
  Nucl.\ Phys.\ B {\bf 635} (2002) 215
 \href{http://arxiv.org/abs/hep-th/0202032}{hep-th/0202032}.


\bibitem{Hadasz:2013bwa}
  L.~Hadasz, M.~Pawelkiewicz and V.~Schomerus,
  ``Self-dual Continuous Series of Representations for $U_q(sl(2))$ and $U_q(osp(1|2))$,''
  JHEP {\bf 1410} (2014) 91
  \href{http://arxiv.org/abs/1305.4596}{arXiv:1305.4596}.


\bibitem{Sarkissian:2009aa}
  G.~Sarkissian,
  ``Defects and Permutation branes in the Liouville field theory,''
  Nucl.\ Phys.\ B {\bf 821} (2009) 607
\href{http://arxiv.org/abs/0903.4422}{arXiv:0903.4422}.

\bibitem{Poghosyan:2015oua}
  H.~Poghosyan and G.~Sarkissian,
  ``On classical and semiclassical properties of the Liouville theory with defects,''
  JHEP {\bf 1511} (2015) 005
  \href{http://arxiv.org/abs/1505.00366}{arXiv:1505.00366}.
  

\bibitem{Fateev:2010za}
  V.~Fateev and S.~Ribault,
  ``Conformal Toda theory with a boundary,''
  JHEP {\bf 1012} (2010) 089
 \href{http://arxiv.org/abs/1007.1293}{arXiv:1007.1293}.    
\bibitem{Fateev:2000ik}
  V.~Fateev, A.~B.~Zamolodchikov and A.~B.~Zamolodchikov,
 ``Boundary Liouville field theory. I: Boundary state and boundary  two-point
  function,''
\href{http://arxiv.org/abs/hep-th/0001012}{arXiv:hep-th/0001012}.
\bibitem{Drukker:2010jp}
  N.~Drukker, D.~Gaiotto and J.~Gomis,
  ``The Virtue of Defects in 4D Gauge Theories and 2D CFTs,''  JHEP {\bf 1106} (2011) 025
    \href{http://arxiv.org/abs/1003.1112}{arXiv:1003.1112}.  
\bibitem{Aguirre:2013zfa}
  A.~R.~Aguirre,
  ``Type-II defects in the super-Liouville theory,''
  J.\ Phys.\ Conf.\ Ser.\  {\bf 474} (2013) 012001
  \href{http://arxiv.org/abs/1312.3463}{arXiv:1312.3463}.
  \bibitem{gam2}
T.~ Shintani, ``On a Kronecker limit formula for real quadratic fields", J.\  Fac.\ Sci.\ Univ.\ Tokyo Sect. 1A Math.
{\bf 24} (1977) 167-199
\bibitem{gam1}
E.~ W.~ Barnes, ``Theory of the double gamma function", Phil.\  Trans.\ Roy.\  Soc {\bf A196} (1901) 265-388
    \end{thebibliography}
\end{document}